\documentclass[12pt]{article}
\usepackage{graphics,color,epsfig,lscape,rotate}
\begin{document}

\title{New Results from CLEO}

\author{D. Cinabro\\
{\small Wayne State University, Department of Physics and Astronomy}\\
{\small Detroit, MI 48201, USA}\\
{\small E-mail: cinabro@physics.wayne.edu}\\
Representing the CLEO Collaboration\\
\\
Plenary Submission to XXXth International Conference on\\
High Energy Physics}

\date{15 September 2000}

\maketitle

\begin{abstract}
The latest results from the CLEO collaboration
are summarized.  An update on the status of the upgraded CLEO~III detector
is also included.
\end{abstract}

\section{Introduction}

	The $\Upsilon (4S)$ has proved to be a rich mine of physics
results.  The study of rare $B$ decays has been a constant challenge
to beyond standard model extensions.  Semileptonic $B$ decays have provided
crucial information on the CKM matrix elements $V_{cb}$ and $V_{ub}$.
Hadronic $B$ decays are also a challenge and provide insight into QCD.
With the beginning of the programs at the asymmetric $B$ factories
at SLAC and KEK the new avenue of CP-violation studies has opened up.
There is also a wealth of physics under the resonance especially in charm
mesons, charm baryons, and taus.  The energy range is also ideally suited
to two photon production studies, and of course the study of the
properties of the $\Upsilon$ and other $b \bar b$ resonances themselves.
CLEO entered into its twentieth year of data taking at the end of
1999.  I summarize here the latest, that is since the 99-00 winter conference
season, results from CLEO.  Even limiting myself to this time slice leaves
no room for tau physics and many results can be only referenced.

	I will first discuss our new results in rare $B$ decays where we
have an unambiguous observation of the gluonic penguin mode, $B \to \phi K$,
new limits on $B$ decays to $\pi^0 \pi^0$, a pair of charged leptons,
$\tau \nu$, and $K \nu \nu$.  Turning to semileptonic $B$ decays I
present our preliminary results 
on $V_{cb}$ in $B \to D^\ast \ell \nu$ decay, and a new measure of
$B$ mixing parameters combining a lepton tag with a partial reconstruction
hadronic tag.  In hadronic $B$ decays we have observed $B \to D^{(\ast)}4\pi$
and studied the resonance substructure of the $4\pi$ system.  We have many
new results on $B \to $ Charmonium including observations of $B \to \eta_c K$
and limits on $\chi_{c0}$ in an attempt to understand our anomalous
results on $B \to \eta$.  Also we test charmonium production models
by observing $B \to \chi_{c1}$ and limiting $\chi_{c2}$.  We have measurements
of $B \to D_s^{(\ast)} D^\ast$ and evidence for $D_s^{(\ast)} D^{\ast \ast}$. 
In the $D_s D^\ast$ decay and $D^\ast 4\pi$ we measure the polarization
of the $D^\ast$ to test a prediction of the factorization ansatz.  We have
observed the first exclusive $B$ decays to nucleons in
$D^{\ast -} p \bar{p} \pi^+$ and $D^{\ast -} p \bar{n}$.
In charm physics we have observed wrong sign $D^0 \to K \pi \pi^0$ and
are working hard on more $D \bar{D}$ mixing and Doubly Cabibbo Suppressed
Decay (DCSD) studies.   In charm baryons we have many new results
including first observations of
the $\Sigma_c^{\ast +}$ and the pair $\Xi^+_{c1}$ and $\Xi^0_{c1}$.  These
complete the set of L = 0 charm baryons, the bulk of which
were first observed by CLEO.  In resonance physics we have a measure of
the rate for the $\Upsilon(4S)$ to decay to charged and neutral $B$ meson
pairs.   We have measurements of $\eta_c$ parameters based on its two
photon production to investigate a puzzle in PQCD.  Finally CLEO
went through a major upgrade to the third major version of our detector.
This began taking physics data in July of 2000, and I will briefly review
the status of CLEO~III.

	The results discussed below are based on the CLEO~II data set
taken with the CLEO~II
detector from 1989 to 1999.  The detector is described 
in detail elsewhere.~\cite{cleo}  About two thirds of the data were taken
from 1995 to 1999 in the CLEO~II.V configuration which replaced the
innermost straw-tube detector of CLEO~II with a high precision silicon
vertex detector.\cite{cleoiiv}  Most of the analyses discussed use
the entire CLEO~II data set with exceptions for systematic error limited
studies and analyses that depend on the precision vertex measuring capabilities
only available in CLEO~II.V.  The total data set has an
integrated luminosity of roughly 14/fb with two thirds taken at
a center of mass energy of about 10.58 GeV on the peak of 
$\Upsilon(4S)$ resonance, corresponding to roughly ten million
$B \bar B$ events, and one third taken at an energy 60~MeV below the
$\Upsilon(4S)$ peak and well below the $B \bar B$ threshold.

\section{Rare $B$ Decays}
\subsection{$B \to \phi K$}

\begin{figure}
\resizebox{6in}{!}
{\includegraphics{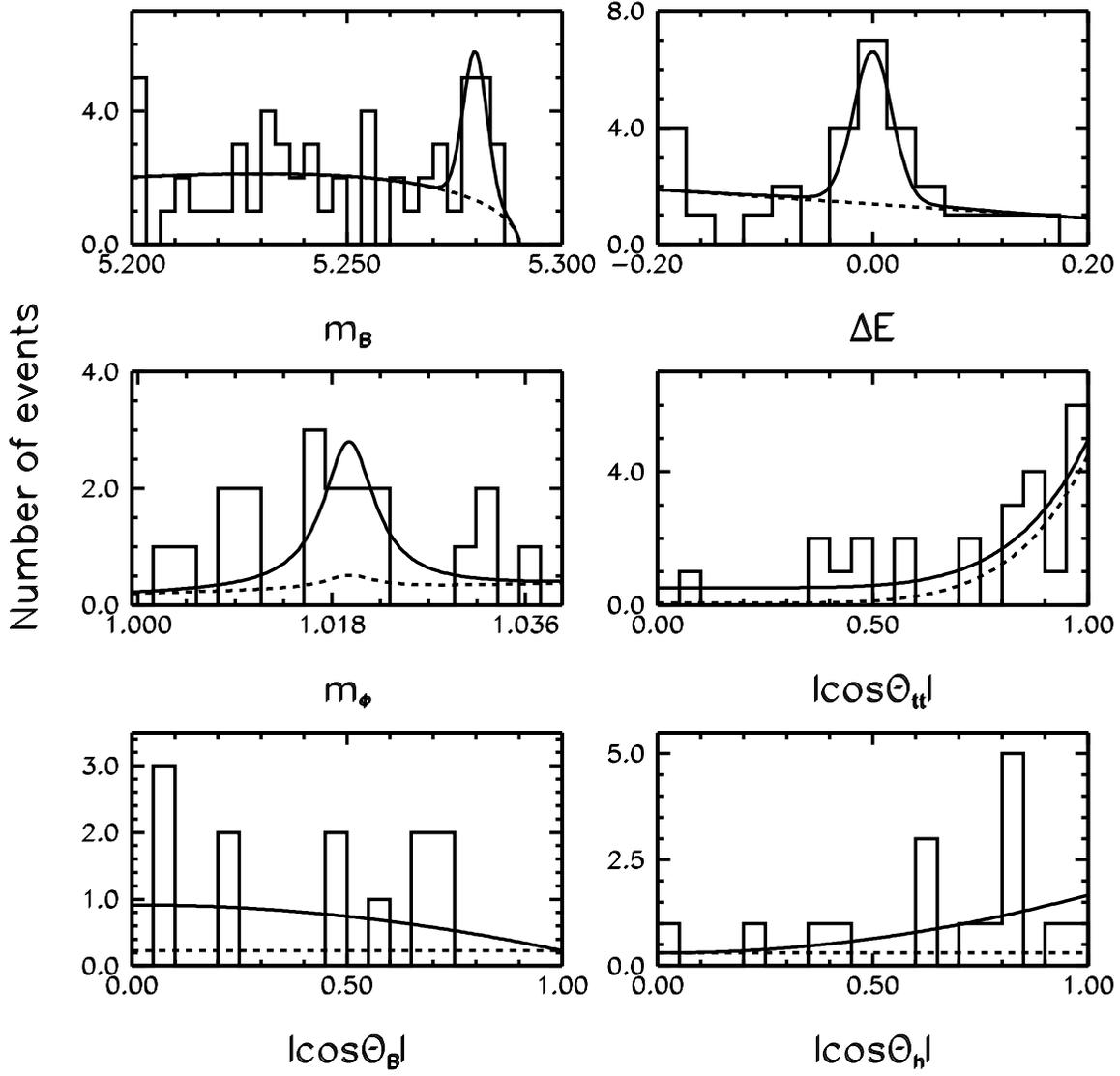}}
\caption{Projections of the $\phi K^-$ data on the six variables
used in the maximum likelihood fit.  The solid lines show the total fit
while the dashed lines show the contribution of the background.}
\label{fig:phikplots}
\end{figure}

	The decay $b \to s g$ produced by the gluonic penguin can be uniquely
tagged when the gluon splits into an $s \bar s$ pair as no other $b$ decay
can produce this final state.  	The mode $B \to \phi K$ is one such
tag of the gluonic penguin and its rate is a vital piece to the rare $B$
decay puzzle.  We search for the signal in both the charged and neutral
modes pairing a reconstructed $\phi \to KK$ candidate with a charged track that
has a specific ionization ($dE/dx$) and time-of-flight consistent
with a kaon or a reconstructed $K^0_s \to \pi\pi$ candidate.  We extract
the yield of signal events by performing an unbinned, maximum likelihood
fit to the six variables shown in Figure~\ref{fig:phikplots} where
the PDF for the signal is taken from simulation and the background
is taken from off resonance data.   The yield, significance, efficiency for 
this procedure and the preliminary branching fractions for the two modes are
displayed in Table~\ref{tab:phik}.
\begin{table}
\caption{Results of likelihood fit.  Note that the efficiencies do
not include any branching fractions}
\begin{center}
\begin{tabular}{|c|c|c|} \hline
 & $B^- \to \phi K^-$ & $B^0 \to \phi K^0_s$ \\ \hline
Signal Yield          & $15.8^{+6.1}_{-5.1}$ & $4.3^{+2.0}_{-2.1}$ \\ \hline
Significance ($\sigma$)& 4.72 & 2.94 \\ \hline
Efficiency (\%) & 49 & 31 \\ \hline
{\cal B} ($\times 10^{-6}$) & $6.4^{+2.5}_{-1.8}$ & $5.9^{+4.0}_{-2.9}$ \\ \hline
\end{tabular}
\end{center}
\label{tab:phik}
\end{table}
Including systematics, we interpret the neutral mode as an upper
limit of  $1.2 \times 10^{-5}$ at the 90\% C.L.,
measure $(6.4^{+2.5+0.5}_{-2.1-2.0}) \times 10^{-6}$ for the charged
mode, and $(6.2^{+2.0+0.7}_{-1.8-1.7}) \times 10^{-6}$ for the average.
The average has a significance of $5.56\sigma$.  The systematics
are dominated by the fit procedure.  All the results are
preliminary.  They agree with theoretical expectations for 
$B \to \phi X_s$\cite{theophik} if the $K$ fraction of $X_s$ is 6-10\%.

\subsection{$B \to \pi^0 \pi^0$}

	Continuing our program of search for all
the $B \to \pi\pi$, $K\pi$,
and $KK$ modes\cite{charmless} to try to gain information on the angles
of the
standard unitarity triangle,\cite{many}  we have a new preliminary
results on the all neutral mode
$B^0 \to \pi^0 \pi^0$.  No signal is observed,
it is expected to be much smaller than our previously observed
$B^0 \to \pi^+ \pi^-$ signal as it is color suppressed, and we set a
90\% C.L. upper limit of $5.7 \times 10^{-6}$ on the branching fraction.
It is interesting to note that in this analysis we have to account
for possible feed down from $B \to \pi\rho$ modes some of 
which we have recently observed.\cite{vector}
  
\subsection{$B \to \ell \ell$}

	Higgs doublet and SUSY extensions to the Standard Model and leptoquark
models predict a large enhancement in the rate of neutral $B$ decays
to two charged leptons.  The leptoquark models can even allow the lepton
flavor violating mode $B^0 \to e^\pm \mu^\mp$ to occur.  We have searched
for these modes and see no evidence for them.\cite{ll}  The results
are summarized in Table~\ref{tab:ll}.
\begin{table}
\caption{Limits on the indicated branching fractions from the CLEO search
for $B^0 \to \ell\ell$ compared with the predicted rates from the
Standard Model.}
\begin{center}
\begin{tabular}{|c|c|c|} \hline
Mode & 90\% C.L.U.L. & Prediction\cite{llpred} \\ \hline
$e^+ e^- $& $8.3\times 10^{-7}$  & $1.9 \times 10^{-15}$ \\ \hline 
$e^\pm \mu^\mp$ & $15\times 10^{-7}$  & 0 \\ \hline 
$\mu^+ \mu^-$ & $6.1\times 10^{-7}$  & $8.0 \times 10^{-11}$ \\ \hline 
\end{tabular}
\end{center}
\label{tab:ll}
\end{table}

\subsection{$B \to \tau\nu$ and $B \to K \nu \nu$}

\begin{figure}
\resizebox{6in}{!}
{\includegraphics{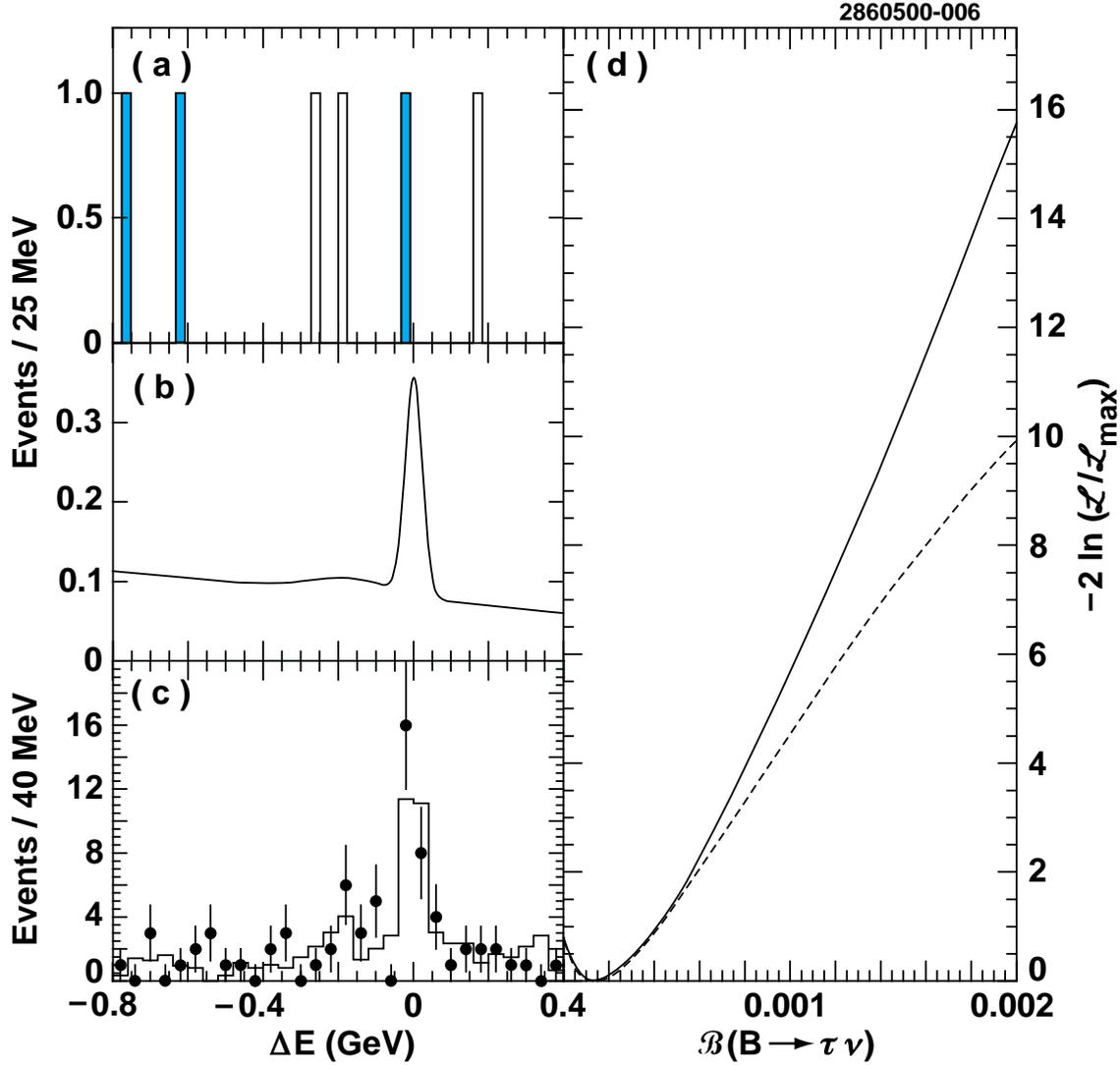}}
\caption{The results of the $B \to \tau \nu$ search.  (a) The six events
that pass all selections.  In the three shaded events the single track
is not consistent with a lepton and these are also candidates for
the $K\nu\nu$ search. (b) The fitting shape used. (c) A comparison
between data and simulation for events opposite tagging $B$'s, but with
two extra charged tracks.  (d) The log likelihood used to calculate
the upper limit.  The solid curve considers statistical errors only;
the dashed includes systematic effects.}
\label{fig:taunu}
\end{figure}

	The charged $B$ can decay via annihilation
to a $W$ of its two internal quarks
to a lepton and neutrino.  The observation and measurement of these decays
are among the most important rare $B$ decays because they provide a
unique constraint on the unitarity triangle when combined with
the $B$ mixing measurements\cite{taunutheo} and provide the cleanest
way to measure the decay constant, $f_B$.   At CLEO we have searched for
the decay $B \to \tau \nu$, the largest mode because of helicity suppression,
by looking for a single track from the $\tau$ decay opposite an inclusively
reconstructed charged $B$ decay to $D$ or $D^\ast$ plus up to four pions
only one of which is allowed to be a $\pi^0$.\cite{taunu}  We look for
eight decay modes of the $D$.
We calibrate this tagging efficiency by running
the analysis on a $B \to D^\ast \ell \nu$ sample and measure the tagging
efficiency with a relative accuracy of 24\% where this error is
dominated by the statistics of the check sample.  The result is shown
in Figure~\ref{fig:taunu}
in terms of the difference between the energy of
the tagging $B$ and the beam energy.
Also shown is the expected distribution from simulated signal events,
and the result of a likelihood fit for the most probable branching fraction.
We obtain an upper limit of $8.4 \times 10^{-4}$ for the branching
fraction at 90\% C.L.  Standard model expectations are in the mid-$10^{-5}$ range.

	This search can be easily modified to look for $B \to K^\pm \nu\nu$
which can be mediated by an electroweak penguin.  Single tracks that
are consistent with leptons are excluded, as shown in Figure~\ref{fig:taunu},
and we obtain an upper limit
on the branching fraction of $2.4 \times 10^{-4}$ at 90\% C.L.

\section{Semileptonic $B$ Decays}
\subsection{$|V_{cb}|$ in $B \to D^\ast \ell \nu$}

	The measurement of $V_{cb}$ is vital to our understanding of the
unitarity triangle as it sets the scale of the entire triangle.  The favorite
technique is to consider the decay $B \to D^\ast \ell \nu$ in the
context of the Heavy Quark Effective Theory (HQET).  The prediction is that
at the kinematic end point where the $D^\ast$ is at rest with respect
to the decaying $B$, $q^2$ is maximal and $w \equiv v_B \cdot v_{D^\ast}$ is 
minimal at 1, the rate for the decay is proportional to $(|V_{cb}|F(1))^2$.
Here $F(w)$ is the universal form factor of HQET.
Thus the strategy of this analysis is measure $d\Gamma/dw$ for
$B \to D^\ast \ell \nu$, extrapolate to the end point, and appeal to
theory to calculate the proportionality to measure $|V_{cb}|$.  In CLEO
this analysis critically depends on the tracking efficiency for the low
momentum pion from the $D^\ast$ decay, is systematically limited mainly
by our ability to measure this, and thus only uses the
first third of the CLEO~II
data set, the data taken in the CLEO~II
configuration.  It is preliminary and is documented more completely 
elsewhere.\cite{dstlnu}

\begin{figure}
\resizebox{6in}{!}
{\includegraphics{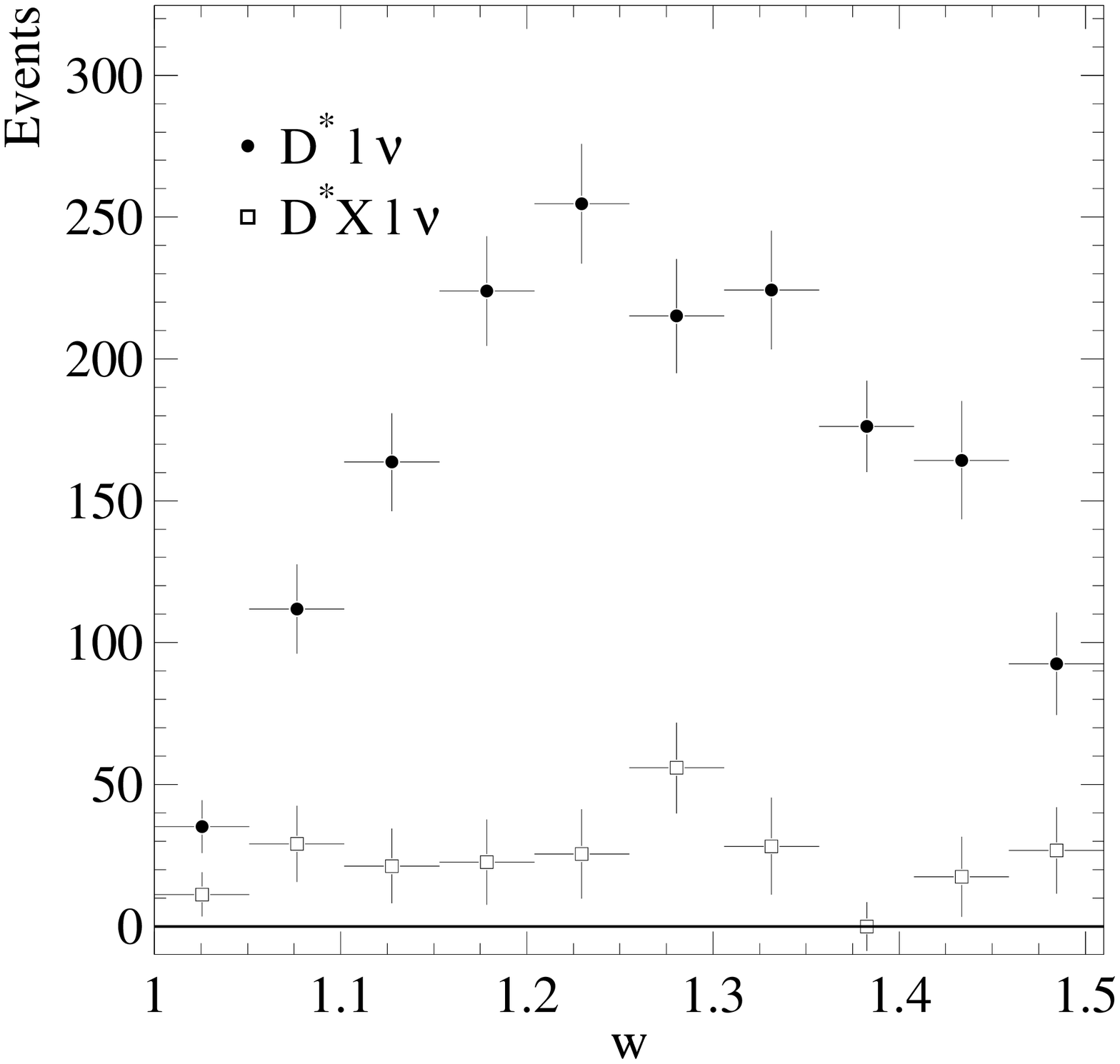}}
\caption{The $D^\ast \ell \nu$ and $D^\ast X \ell \nu$ yields in each $w$ bin.}
\label{fig:wraw}
\end{figure}

	The $d\Gamma/dw$ distribution is measured by fitting the
distribution of $\cos\theta_{B-D^\ast \ell}$ which can be computed from
observed $D^\ast$-$\ell$ pairs and the known beam energy and $B$ mass
assuming the only missing particle from the $B$ decay is a massless
neutrino.  Backgrounds are determined from the data from non-$B \bar B$
events using off resonance data and combinatorics from $D^\ast$ sidebands.
We use the simulation to model correlated backgrounds, where both the 
$D^\ast$ and $\ell$ are from the same $B$ decay, uncorrelated backgrounds
where they are from different $B$ decays, and we ignore the
small contribution from fakes.  Combinatorics
are the largest background source, about 6\%, 
continuum and uncorrelated are about 4\% each, and correlated background
is about 0.5\%.  The result is shown in Figure~\ref{fig:wraw}.  There
remains a background of $D^\ast X \ell \nu$ with contributions from 
$B$ semileptonic decay to the so called $D^{\ast\ast}$ resonances and
to non-resonant $D^\ast$ plus at least one pion.  These are modeled with
the simulation and generously varied to test their systematic effect.

\begin{figure}
\resizebox{6in}{!}
{\includegraphics{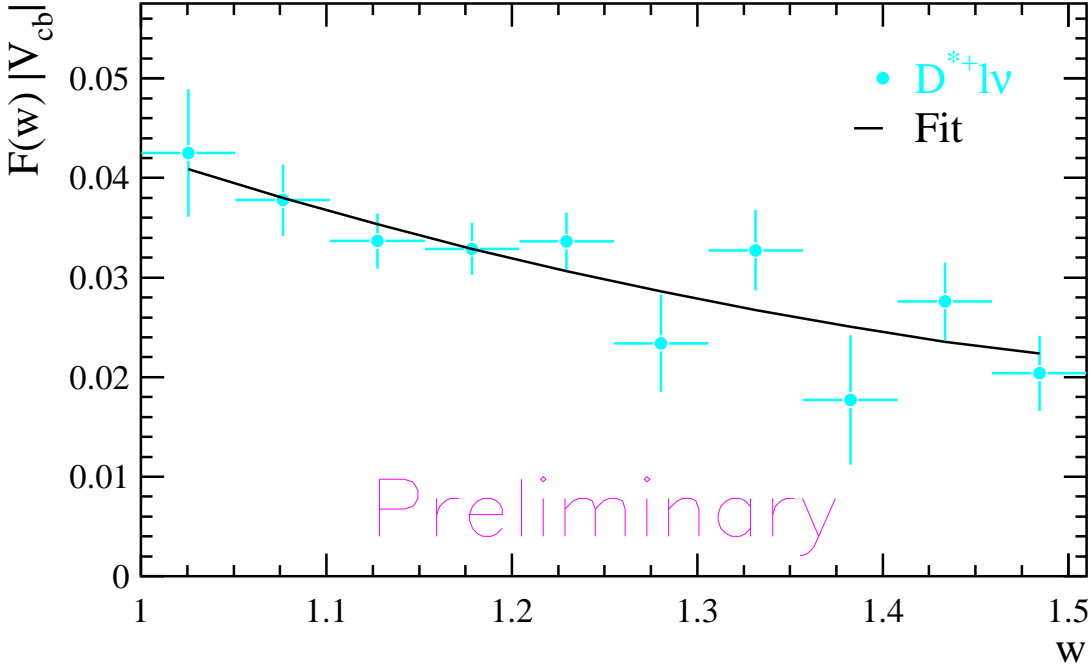}}
\caption{The result of the fit to the corrected $d\Gamma/dw$ distribution.}
\label{fig:wfit}
\end{figure}

	The partial width is given by\cite{richburch}
\begin{equation}
\frac{d\Gamma}{dw}  =  \frac{G_F^2 |V_{cb}|^2}{48 \pi^3}G(w)F(w)^2
\label{eq:w}
\end{equation}
where $G(w)$ is a known kinematic function.  The universal HQET
form factor $F(w)$ depends on two form factor ratios $R_1(w)$
and $R_2(w)$, and a normalization $h_{A_1}(w)$.  These can be constrained
with dispersion relations.\cite{cap}
The dependence can then be reduced to a ``slope,'' $\rho^2$, and $R_1(1)$,
and $R_2(1)$.  We use our
measured values for $R_1(1)$ and $R_2(1)$.\cite{r12}
To extract the intercept we start with Figure~\ref{fig:wraw} subtract the
remaining background, correct for efficiency and resolution, and
fit to Equation~\ref{eq:w} with the constraints discussed above.
The result is shown in Figure~\ref{fig:wfit}.

	Systematic errors are dominated by the uncertainty in
the slow pion detection efficiency as a function of momentum.
The combinatoric background and the lepton ID efficiency are also
important effects.  For the
slope the uncertainties on $R_1(1)$ and $R_2(1)$
are the largest effect and the branching fraction has significant
uncertainty from the measurements of the $D^\ast \to D\pi$ and
the lepton ID efficiency.  We find
\begin{equation}
|V_{cb}|h_{A_1}(1) = 0.0424 \pm 0.0018 \pm 0.0019,
\end{equation}
\begin{equation}
\rho^2             = 1.67   \pm 0.11 \pm 0.22,
\end{equation}
\begin{equation}
{\cal B}(\bar{B}^0\to D^{\ast+}\ell^-\bar{\nu}) = (5.66 \pm 0.29 \pm0.33)\%,
\end{equation}
with a correlation coefficient of 0.90 between $|V_{cb}|h_{A_1}(1)$
and $\rho^2$.
This slope is defined differently than our previous analysis which
assumed a linear dependence.  If we use a linear fit we obtain
consistent results.
Using $h_{A_1}(1) = 0.913 \pm 0.042$\cite{babar} this gives
\begin{equation}
|V_{cb}| = 0.0464\pm0.0020 \pm0.0021 \pm0.0021
\end{equation}
where the last error is due to the uncertainty on $h_{A_1}(1)$.
This result is somewhat higher than our previous result, but
this is partly due to the different assumption made about the
shape of $d\Gamma/dw$.  If the same assumption is made we obtain
consistent results.
 
\subsection{$B^0$ Mixing}

	The measurement of $B_d$-mixing is an important cross check
and provides a valuable input to the extraction CP-violation
parameters in the $b$ system.  We have used a new technique
at the $\Upsilon(4S)$\cite{mixing} which combines the traditional
lepton tag with a partial reconstruction of $\bar{B^0} \to D^{\ast+} \pi^-$
or $\rho^-$.  This partial reconstruction technique only observes
the fast $\pi^-$ or $\rho^-$ from the $B$ decay and the slow pion
from the $D^\ast$ decay.  This increases the statistics
over the
dilepton method and reduces the dilution due to charged $B$ contamination.
Also the partial reconstruction is very clean.  The complete
analysis has statistics of about 2000 doubly tagged events with
a dilution of 13\% and only 3\% mistagging.  This leads to the
best single measure for the probability of $B_d$ mixing
of $\chi_d = 0.198 \pm 0.013 \pm 0.014$.  Note that this
measurement has very different sources of systematic errors
that the lifetime based measurements from LEP and the asymmetric
$B$ factories.  We can also do the analysis
comparing $B^0 B^0$ events versus $\bar{B^0}\bar{B^0}$ and obtain
limits on $|\Re (\epsilon_B)| < 0.034$, the $B$ system analog
of the $\epsilon$ parameter in the $K$ system,
and combined with LEP measures of $\Delta m_d$ and lifetime
$\Delta \Gamma_d/2\Gamma_d = |y_d| < 0.41$, both at the 95\% C.L.
This is the first non-trivial limit on $y_d$. 

\section{Hadronic $B$ Decays}
\subsection{$B \to D^{(\ast)}4\pi$}

\begin{figure}
\resizebox{6in}{!}
{\includegraphics{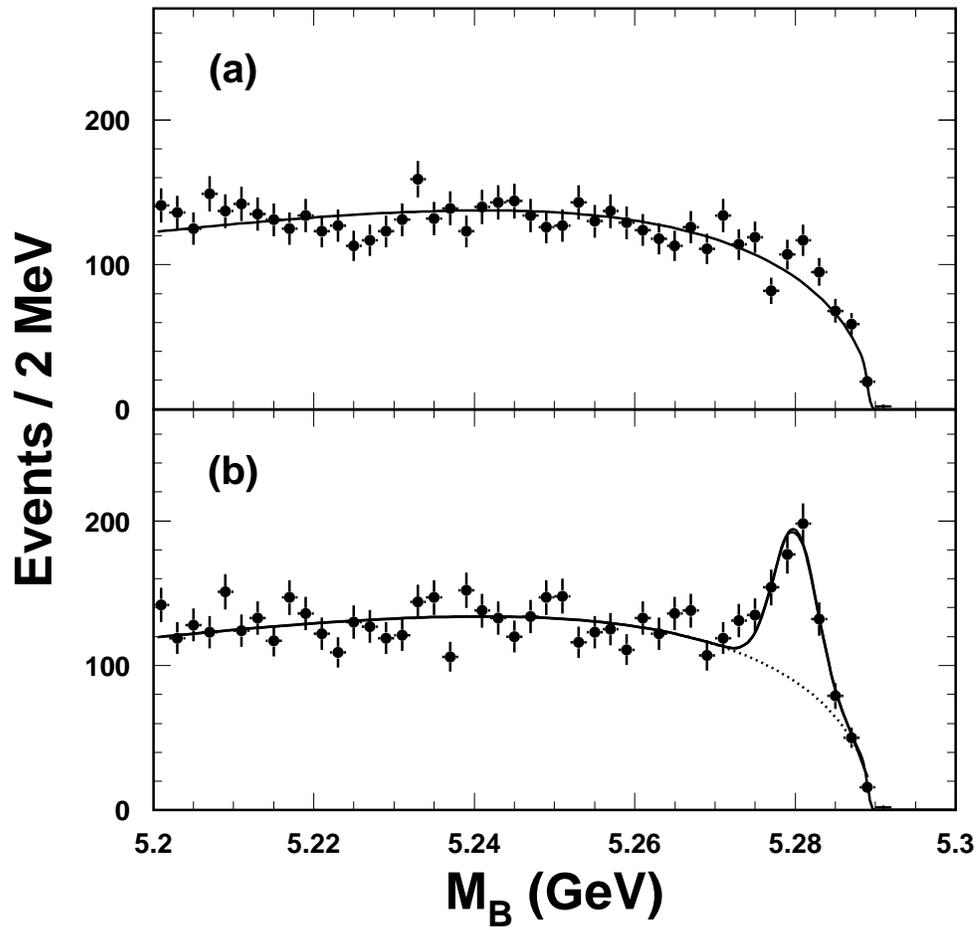}}
\caption{The beam constrained mass for $B \to D^{\ast+}\pi^+\pi^-\pi^-\pi^0$
with $D^0 \to K\pi$.  The top distribution is for a $\Delta E$ sideband,
and the bottom is for $\Delta E$ consistent with zero.}
\label{fig:D4pimass}
\end{figure}

	Despite the large progress in the understanding of hadronic
$B$ decays, only a small fraction of the hadronic branching ratio
has been measured.  The majority of the measured modes are low multiplicity,
and thus we are motivated to search for higher multiplicity modes.
We investigate $B \to D^{(\ast)}4\pi$\cite{d4pi} as shown in
Figure~\ref{fig:D4pimass} and see a clear signal.

\begin{figure}
\resizebox{6in}{!}
{\includegraphics{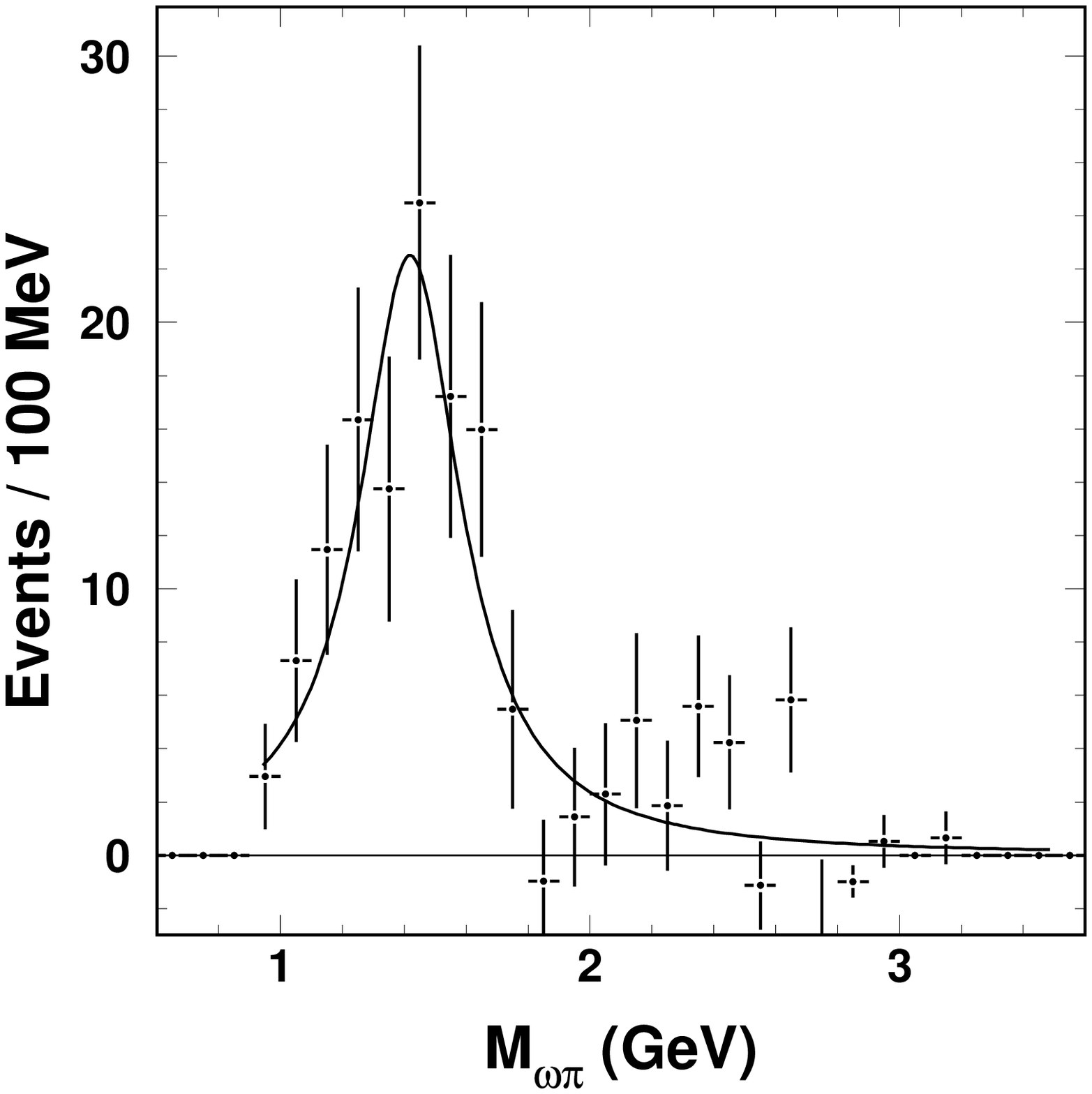}}
\caption{The invariant mass spectrum of the $\omega \pi^-$ for the
final state $D^{\ast+}\pi^+\pi^-\pi^-\pi^0$ combining all $D$ decay
modes.  This is the spectrum determined from fitting the yield in
the beam constrained mass distribution and displaying a fit
to Breit-Wigner function.}
\label{fig:omegapi}
\end{figure}

	The decays $B \to D^{(\ast)}(2-3)\pi$ are dominated by resonant decays
to the $\rho$ and the $a_0$.  This motivates a search for substructure
in the $4\pi$ system.  We do see a clear signal of an $\omega$ in
the $2\pi\pi^0$ system.  Combining the $\omega$ with the remaining
charged $\pi$ we obtain Figure~\ref{fig:omegapi}.
We clearly see a resonance and determine it to have a mass
of $1418 \pm 26 \pm 19$ MeV and width of $388 \pm 41 \pm 32$ MeV.
By studying the
angular distribution of the decays to this resonance we find that it
has $J^P = 1^-$.  We identify this resonance as the $\rho'$.  This
$\rho'$ mass and width measurement are the most accurate and have little 
model dependence.

\begin{figure}
\resizebox{6in}{!}
{\includegraphics{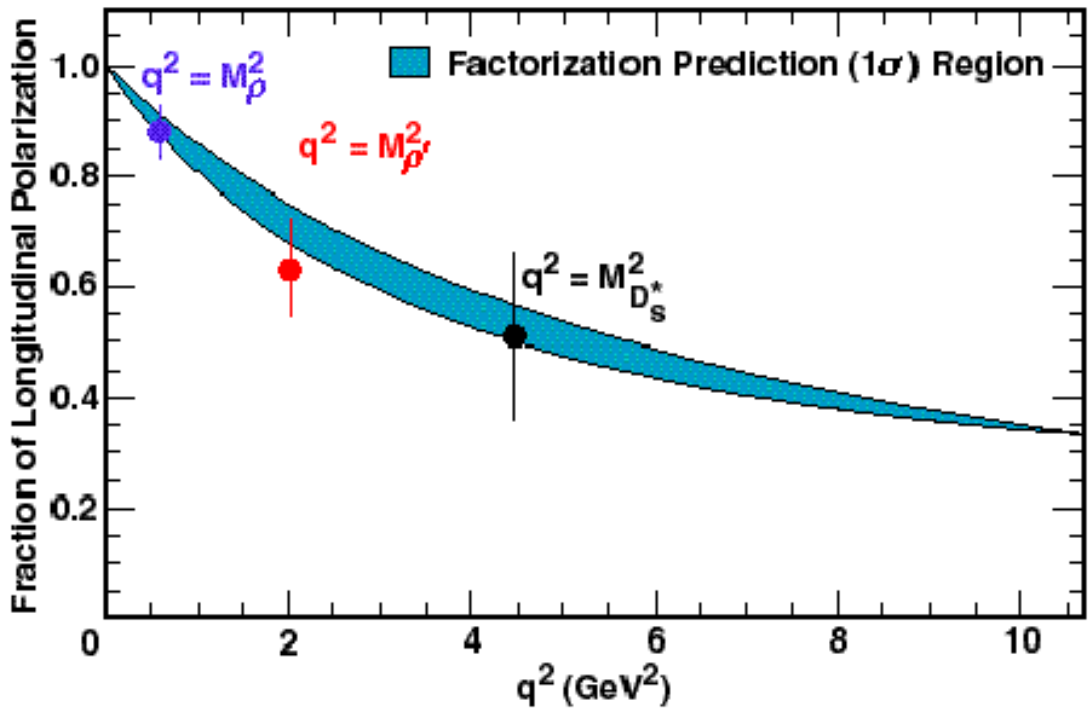}}
\caption{The fraction of $D^\ast$ longitudinal polarization for three
hadronic $B$ decays.  The measurements are compared with the prediction
of the factorization model.}
\label{fig:longpol}
\end{figure}

	Table~\ref{tab:d4pi} summarizes our observations in this mode.
\begin{table}
\caption{Measured branching fractions.}
\begin{center}
\begin{tabular}{|c|c|} \hline
Mode & $\cal{B}$(\%) \\ \hline
$\bar{B^0} \to D^{\ast +}\pi^+\pi^-\pi^-\pi^0$   & $1.72\pm 0.14\pm0.24$ \\ \hline 
$\bar{B^0} \to D^{\ast +}\omega\pi^-$            & $0.29\pm 0.03\pm0.04$ \\ \hline
$\bar{B^0} \to D^{+}\omega\pi^-$                 & $0.28\pm 0.05\pm0.03$ \\ \hline
$B^- \to D^{\ast 0}\pi^+\pi^-\pi^-\pi^0$   & $1.80\pm 0.24\pm0.25$ \\ \hline 
$B^- \to D^{\ast 0}\omega\pi^-$            & $0.45\pm 0.10\pm0.07$ \\ \hline 
$B^- \to D^{0}\omega\pi^-$                 & $0.41\pm 0.07\pm0.04$ \\ \hline 
\end{tabular}
\end{center}
\label{tab:d4pi}
\end{table}
The $\rho'$ saturates the $\omega\pi$ final
states.  In the $D^\ast \rho'$
modes we have measured the longitudinal polarization of the $D^\ast$ to
be $63 \pm 9$\%.  This can be compared to a prediction of the factorization
model that this polarization should be the same as in the semileptonic
decay $B \to D^\ast \ell \nu$ at $q^2$ equal to the mass of the $\rho'$.
The comparison of this measurement with this prediction of factorization
is shown in Figure~\ref{fig:longpol}.
The $D^\ast \rho$ measure is from a previous CLEO result\cite{drho} and the
$D^\ast D_s$ result is discussed below.

\subsection{$B \to $ Charmonium}

	We have many new results on $B$ decays to charmonium.
Motivated by our observation of surprisingly large $B$ decay rates
to $\eta' X$\cite{eta} we search for $B$ decays
to $\eta_c K$ and with trivial modifications to $\chi_{c0} K$.\cite{etac}
If as has been suggested that the large $\eta' X$ rate is due to
intrinsic charm content of the $\eta'$ then we expect an enhancement
in the $\eta_c K$ rate.  Our observations are summarized in 
Table~\ref{tab:etac}.
\begin{table}
\caption{Measured branching fractions or upper limits.  The third error
is due to $\eta_c$ branching fractions.}
\begin{center}
\begin{tabular}{|c|c|} \hline
Mode & $\cal{B}$ or 90\% C.L.U.L($\times 10^{-3}$) \\ \hline
$B^+ \to \eta_c K^+$    & $0.69\pm 0.24\pm 0.08\pm 0.20$ \\ \hline
$B^0 \to \eta_c K^0$    & $1.09\pm 0.49\pm 0.12\pm 0.31$ \\ \hline
$B^+ \to \chi_{c0} K^+$ & $<0.48$ \\ \hline 
$B^0 \to \chi_{c0} K^0$ & $<0.50$ \\ \hline 
\end{tabular}
\end{center}
\label{tab:etac}
\end{table}
The rates for $\eta_c K$ are similar to those for $J/\psi$ indicating
no unexpected enhancement.  Invoking factorization we can
measure $f_{\eta_c} = 335 \pm 75$ MeV also agreeing with expectation.
It seems that the charm content of the $\eta'$ is not the explanation
for the anomalous $B \to \eta' X$ decay rates.

\begin{figure}
\begin{center}
\begin{tabular}{cc}
\resizebox{3in}{!} {\includegraphics{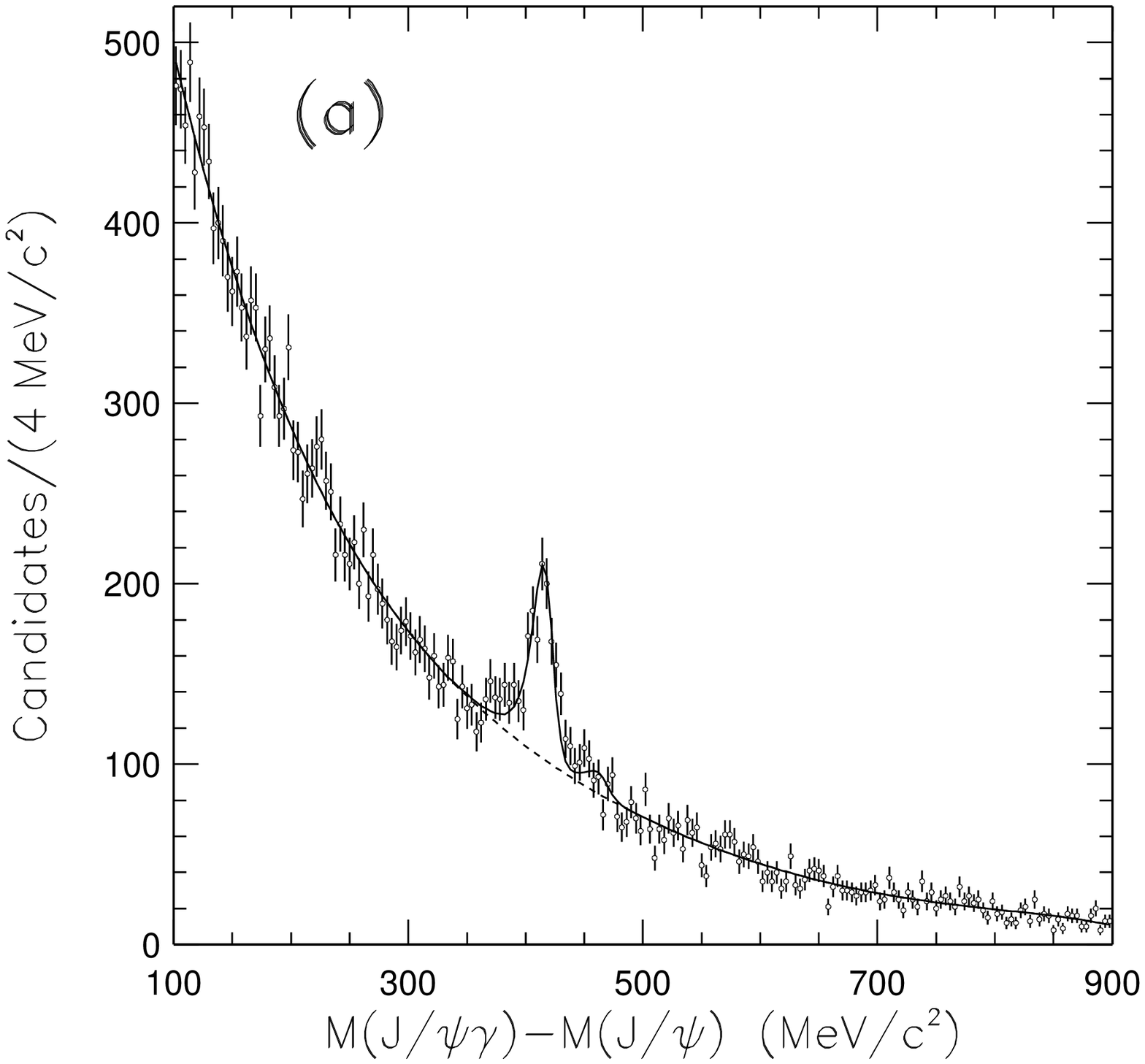}} &
\resizebox{3in}{!}{\includegraphics{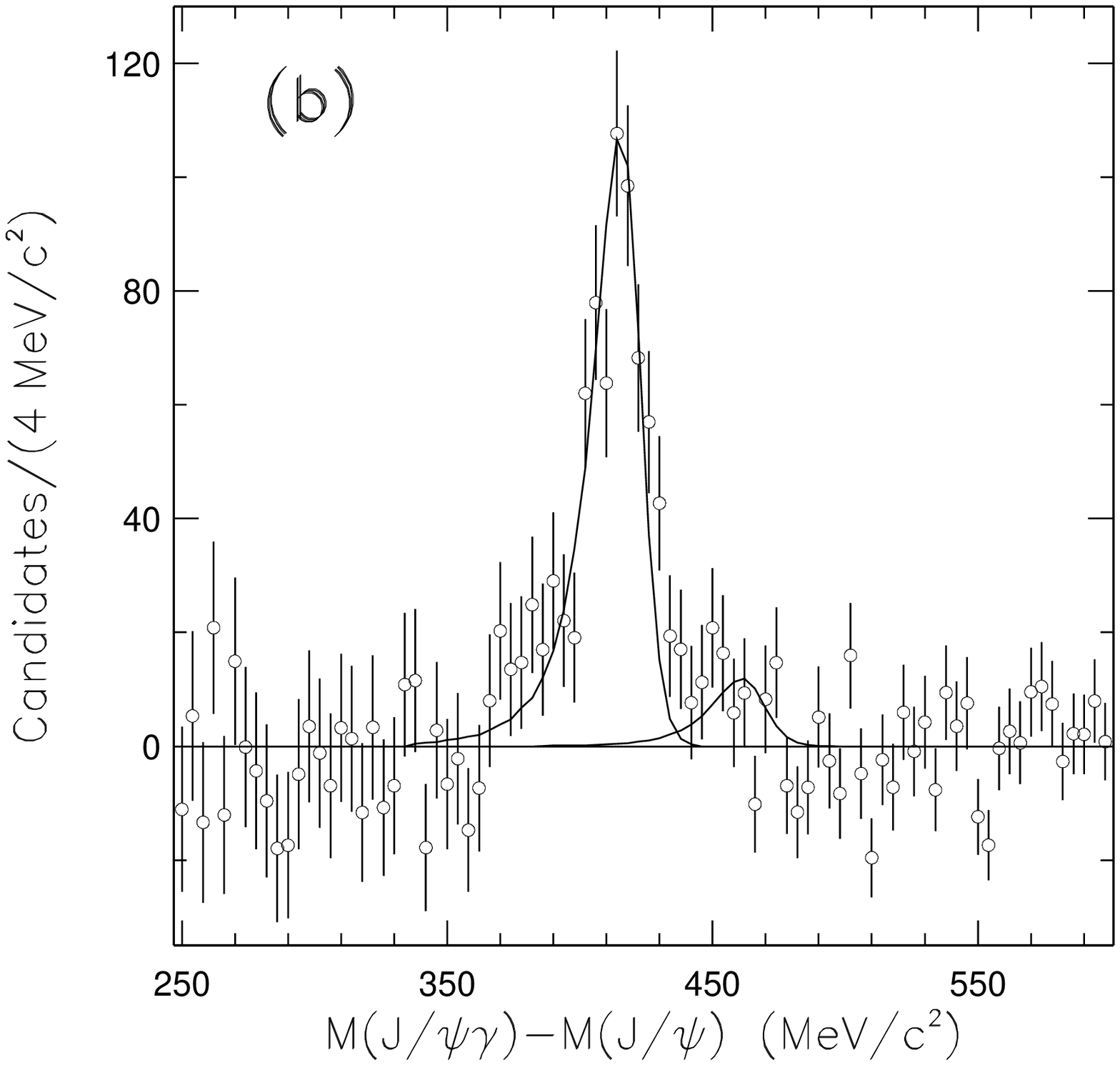}}
\end{tabular}
\end{center}
\caption{The $M(J/\psi \gamma) - M(J/\psi)$ distribution.  The plot
on the right
subtracts off the background fit displayed in the plot on the left.}
\label{fig:chic1}
\end{figure}

	The measurement of a large rate for high-$P_T$ charmonium
at the Tevatron is a challenge to theoretical models of charmonium
production.\cite{octet}  An especially clean test is provided by
measuring the $\chi_{c2}$-to-$\chi_{c1}$ production ratio in $B$
decays.\cite{c1}
Figure~\ref{fig:chic1} shows our basic result.
We do not see significant $\chi_{c2}$ production, and our results
are summarized in Table~\ref{tab:chic1}.
\begin{table}
\caption{Measured branching fractions.  Direct
means that feed down from $B \to \psi(2S)$ has been subtracted }
\begin{center}
\begin{tabular}{|c|c|} \hline
Mode & $\cal{B}$($\times 10^{-3}$) \\ \hline
$B^0 \to \chi_{c1}X)$             & $4.14 \pm 0.31 \pm 0.40$ \\ \hline 
$B^0 \to \chi_{c1}[\rm direct]X)$ & $3.83 \pm 0.31 \pm 0.40$ \\ \hline
\end{tabular}
\end{center}
\label{tab:chic1}
\end{table}
We also limit ${\cal B}(B^0 \to \chi_{c2}[\rm direct]X)/
          {\cal B}(B^0 \to \chi_{c1}[\rm direct]X) < 0.44$ at the
95\% C.L.
The results are preliminary and do not support the prediction
of the color-octet model that the $\chi_{c2}$-to-$\chi_{c1}$ production ratio
should be larger than 0.5.

	We have new measurement of the exclusive two body 
$B$ decays to charmonium that are being used in CP violation measurements
at the asymmetric $B$ factories:
${\cal B}(B \to J/\psi K^0) = (9.5\pm0.8 \pm 0.6)\times 10^{-4}$;
${\cal B}(B \to J/\psi \pi^0) = (2.5\pm1.0 \pm 0.2)\times 10^{-5}$;
and ${\cal B}(B \to \chi_{c1} K^0) = (3.9\pm1.6 \pm 0.4)\times 10^{-4}$.\cite{twobodycc}
We have searches
for direct CP violation in charged $B$ decays to charmonium:
${\cal A}_{CP}(B^\pm \to J/\psi K^\pm) = (+1.8 \pm 4.3 \pm 0.4)$
and ${\cal A}_{CP}(B^\pm \to J/\psi (2S) K^\pm) = (+2.0 \pm 9.1 \pm 1.0)$.\cite{acpcharm}
We also have precision measurements of the $B$ meson masses in
$B \to \psi^{(\prime)} K$: $m(B^0) = 5279.1 \pm 0.7 \pm 0.3$ MeV and
$m(B^+) = 5279.1 \pm 0.4 \pm 0.4$ MeV.\cite{mass}

\subsection{$B \to D_s^{(\ast)}D^\ast$}

	There is a substantial discrepancy between the inclusive and sum of
the exclusive $B$ decays to $D_s$.  We have used a new technique
where a fully reconstructed $D_s \to \phi \pi$ is combined
with a partially reconstructed $D^\ast$, whose soft pion is the only
observed decay product, to measure exclusive $B$ decays to $D_s$.\cite{ds}
This partial reconstruction technique results in much higher statistics
than previous analysis, and much more accurate measurements can be made.
We observe evidence for $B \to D_s^{(\ast)}D^{\ast \ast 0}$ decays.
The results are summarized in Table~\ref{tab:ds}.
\begin{table}
\caption{Measured branching fractions.  The third error is due
to the $D_s \to \phi \pi$ branching fraction.}
\begin{center}
\begin{tabular}{|c|c|} \hline
Mode & $\cal{B}$(\%) \\ \hline
$B^0 \to D_s D^{\ast -}$       & $1.01\pm 0.18\pm 0.10\pm 0.28$ \\ \hline 
$B^0 \to D_s^\ast D^{\ast -}$  & $1.82\pm 0.37\pm 0.24\pm 0.46$ \\ \hline 
$B^+ \to D_s^{(\ast)} D^{\ast \ast 0}$ & $2.73\pm 0.78\pm 0.48\pm 0.68$ \\ \hline 
\end{tabular}
\end{center}
\label{tab:ds}
\end{table}
We also measure the longitudinal polarization of the $D_s^{\ast} D^{\ast -}$
production to be $(51 \pm  14 \pm 4)$\%.  This is compared
with the prediction of factorization in Figure~\ref{fig:longpol},
and agrees well.

\subsection{$B \to $ Nucleons}

\begin{figure}
\begin{center}
\begin{tabular}{cc}
\resizebox{3in}{!} {\includegraphics{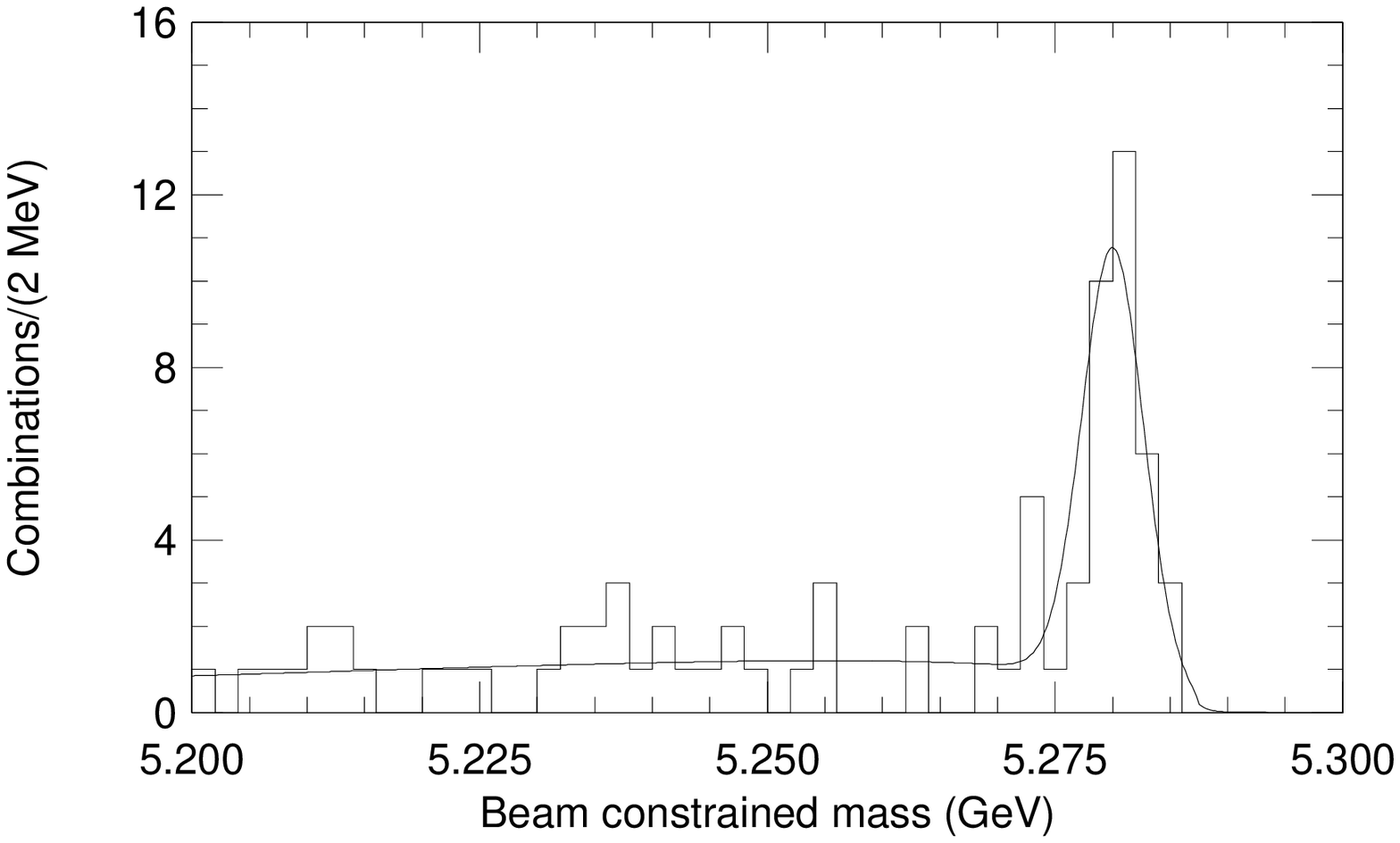}} &
\resizebox{3in}{!}{\includegraphics{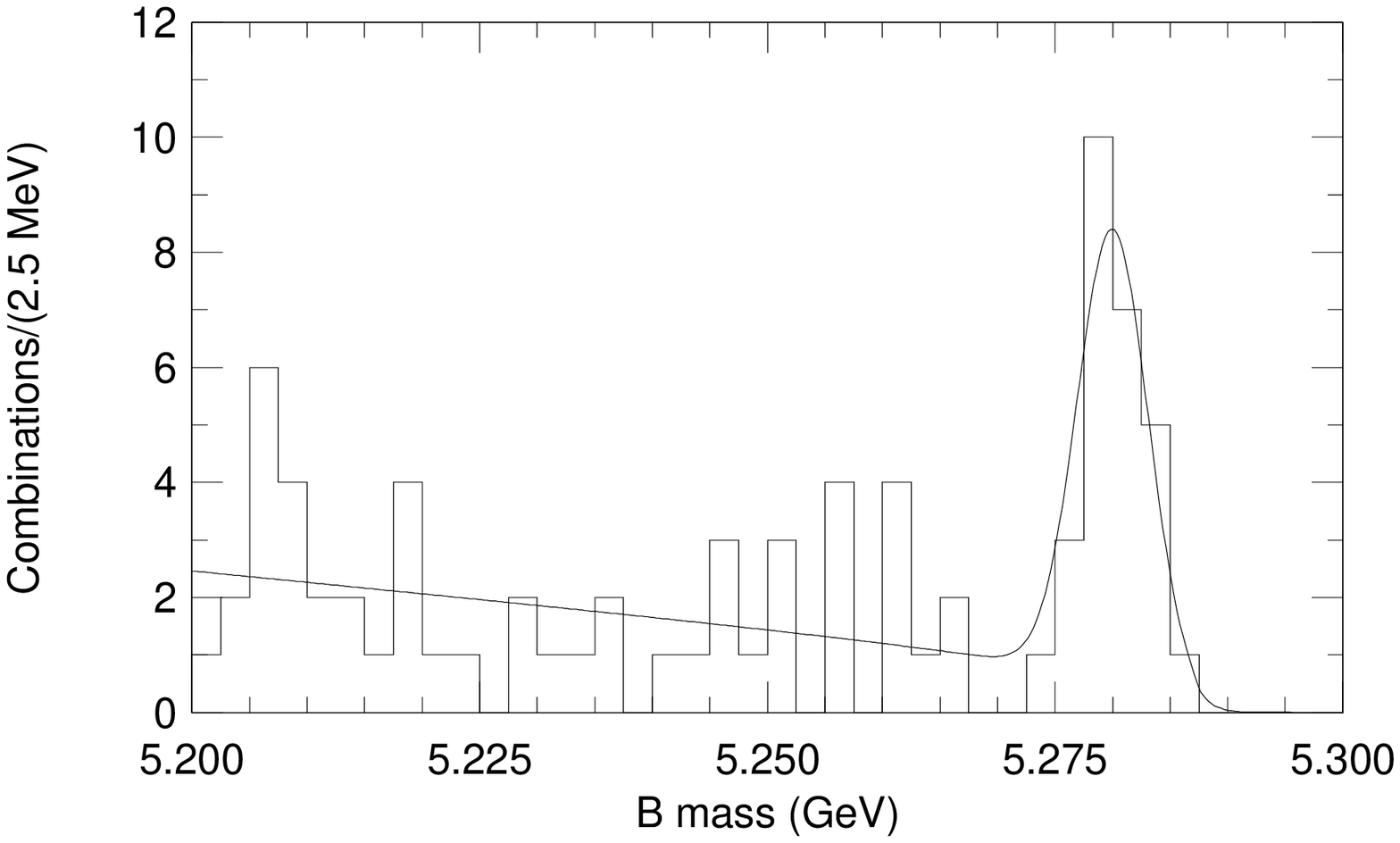}}
\end{tabular}
\end{center}
\caption{The beam constrained mass for the $B \to D N\bar{N}X$ signals
On the left is $D^{\ast -}p \bar{p} \pi^+$ and on the
right is $D^{\ast -}p \bar{n}$.}
\label{fig:bary}
\end{figure}

	A unique feature of the $B$ meson is that the large mass
of the $b$ quark allows weak decays to baryon-anti-baryon pairs.
Measurements of the inclusive rate of $B \to \Lambda_c X$ lead to
estimates that the $B \to \bar{\Lambda}_c N X$, where $N$ is a proton
or neutron, rate accounts for
only half of the baryon production observed in $B$ decays.
We are thus motivated to search for $B \to D N \bar{N} X$.\cite{bary}  We
do observe such decays in the modes and with rates given in
Table~\ref{tab:bary}, and the signals are displayed in
\begin{table}
\caption{Measured branching fractions.}
\begin{center}
\begin{tabular}{|c|c|} \hline
Mode & ${\cal B}(\times 10^{-4})$ \\ \hline
$B^0 \to D^{\ast -}p \bar{p} \pi^+$ & $ 6.6 \pm 1.4 \pm 1.0$ \\ \hline 
$B^0 \to D^{\ast -}p \bar{n}$       & $14.5 \pm 3.2 \pm 2.7$ \\ \hline 
\end{tabular}
\end{center}
\label{tab:bary}
\end{table}
Figure~\ref{fig:bary}.  Note that the $B^0 \to D^{\ast -} p \bar{n}$
mode is observed via the annihilation of the $\bar{n}$ in our
calorimeter.  We are not sensitive to the charge conjugated mode
with a $n$.  These measurements account
for a substantial fraction of the non-$\Lambda_c$ $B$ to baryon
decay rate. 

\section{Charm Physics}
\subsection{$D^0$ Mixing and Doubly Cabibbo Suppressed Decays}

	One of the most exciting areas in the last year has been
new probes with a factor of three more sensitivity for
$D$-mixing and Doubly Cabibbo Suppressed Decays (DCSD).\cite{dmix}
We are working hard at CLEO to make use of our clean event environment
to search in many $D$ decay modes to improve on our results.  All of
these analyses are only done with the superior vertex resolution of
the CLEO II.V detector.

\begin{figure}
\begin{center}
\begin{tabular}{cc}
\resizebox{3in}{!} {\includegraphics{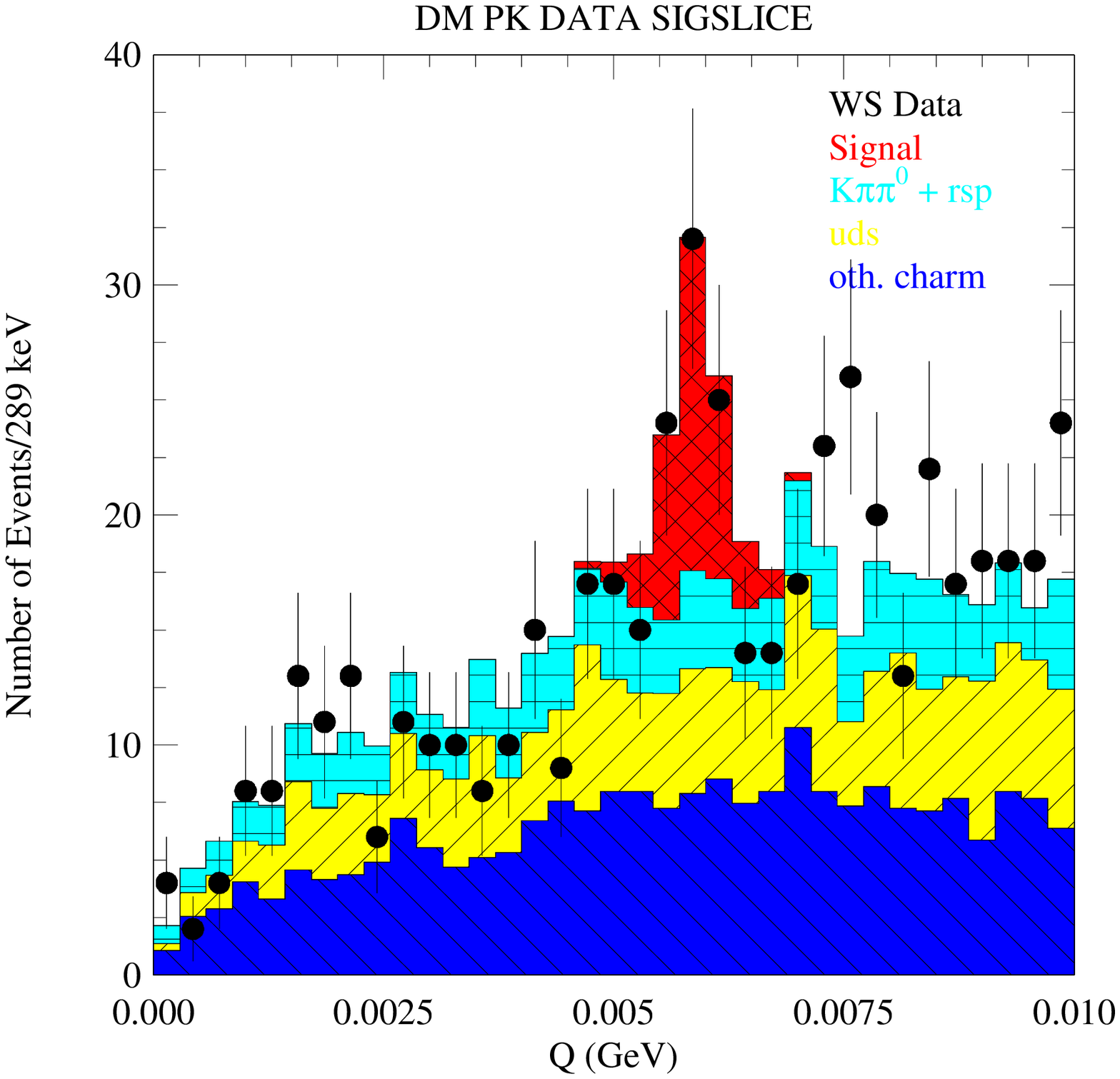}} &
\resizebox{3in}{!} {\includegraphics{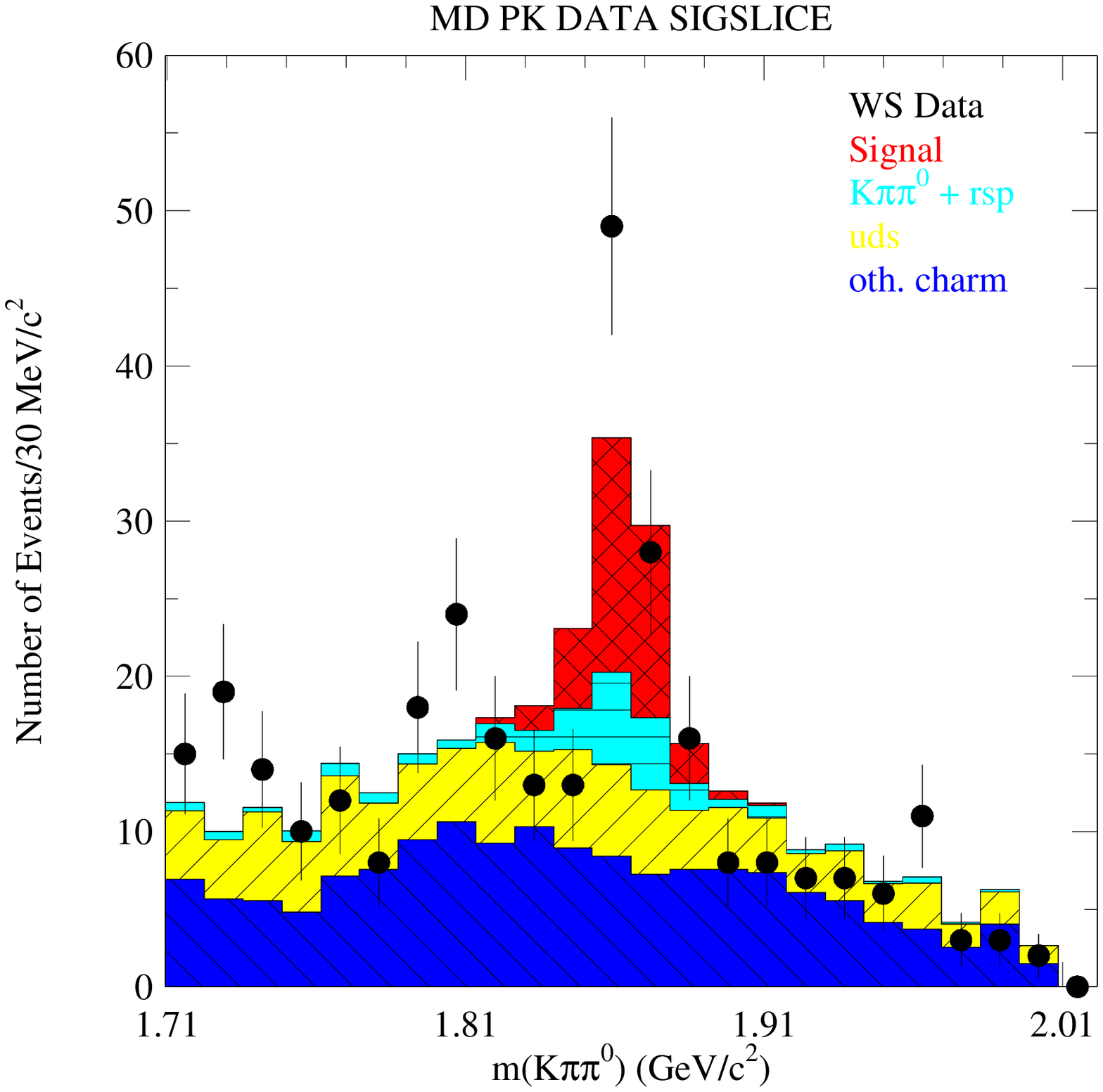}}
\end{tabular}
\end{center}
\caption{The wrong sign $D \to K \pi \pi^0$ signal.  On the right
is the $Q$ distribution and on the left is the $K\pi\pi^0$ invariant
mass.}
\label{fig:kpipi0}
\end{figure}

\begin{figure}
\resizebox{6in}{!}
{\includegraphics{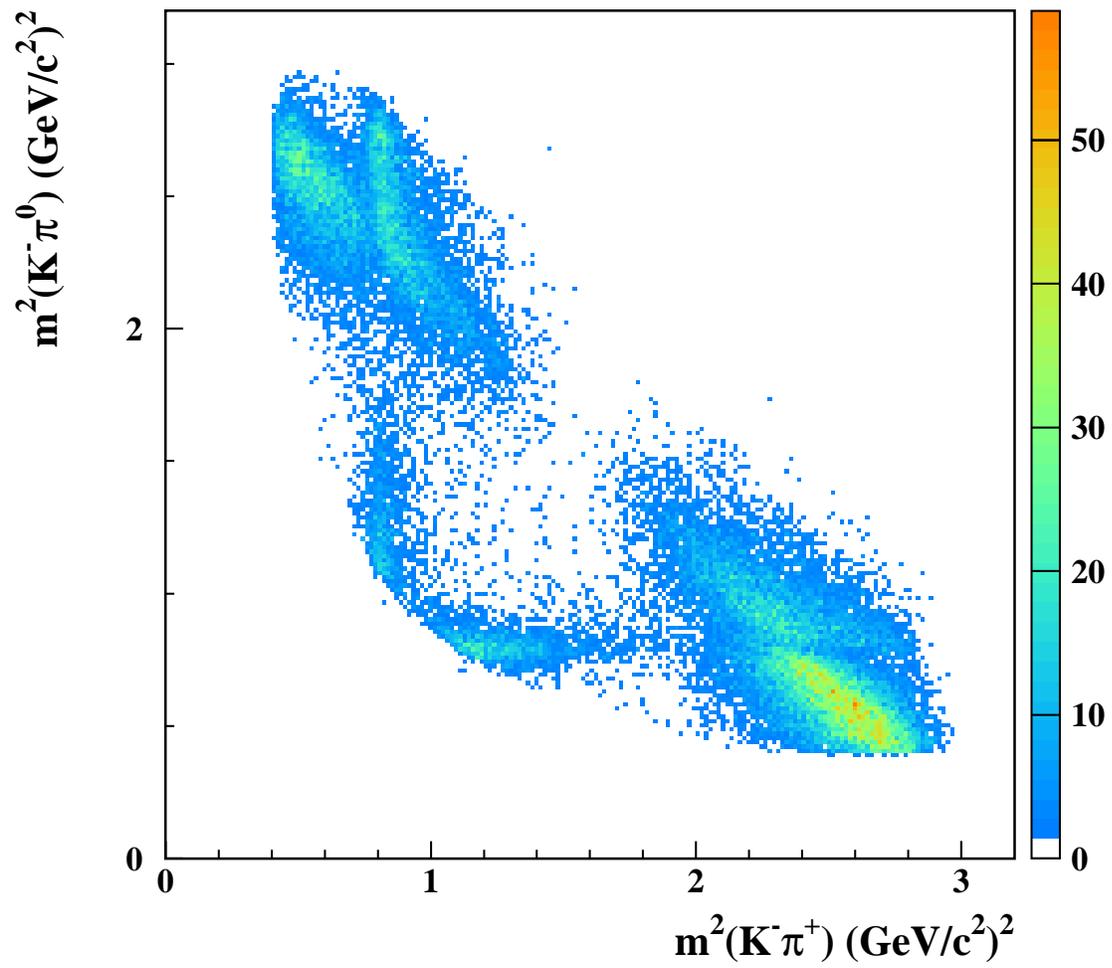}}
\caption{The Dalitz plot for right sign $D \to K\pi\pi^0$ decay
in the CLEO data.}
\label{fig:kpipi0dal}
\end{figure}

	The latest work is preliminary.  We do have
a significant wrong sign signal in $D \to K \pi \pi^0$ decay,
shown in Figure~\ref{fig:kpipi0}, of $39 \pm 10 \pm 7$ events
compared to a right sign yield of over 9000.
We are working hard to turn this yield into a rate, but
the Dalitz structure of the wrong sign decay may be different
than the right sign decay leading to a difference in efficiency
for mixed and DCSD wrong sign events.  The resonance substructure
for $D \to K \pi \pi^0$ is very rich with three dominant
modes
clearly visible in the Dalitz plot for the right sign signal
shown in Figure~\ref{fig:kpipi0dal}, clear signs of interference,
and many other smaller amplitudes that can contribute.

	We are also working on the lifetime analyses of the $K \pi \pi^0$
mode; the CP even eigenstates $KK$ and $\pi\pi$; CP odd eigenstates
$K_s^0 \phi$, $K_s^0 \rho^0$, and  $K_s^0 \omega$; and in the
the semileptonic decays $K \ell \nu$ and $K^\ast \ell \nu$.  One of
the first steps on this road is a preliminary measure of the 
CP asymmetries ${\cal A}_{CP}(D^0 \to KK) = (+0.04 \pm 2.18 \pm 0.84)\%$
and ${\cal A}_{CP}(D^0 \to \pi\pi) = (+1.94 \pm 3.22 \pm 0.84)\%$.

\subsection{Charm Baryons}

\begin{figure}
\resizebox{6in}{!}
{\includegraphics{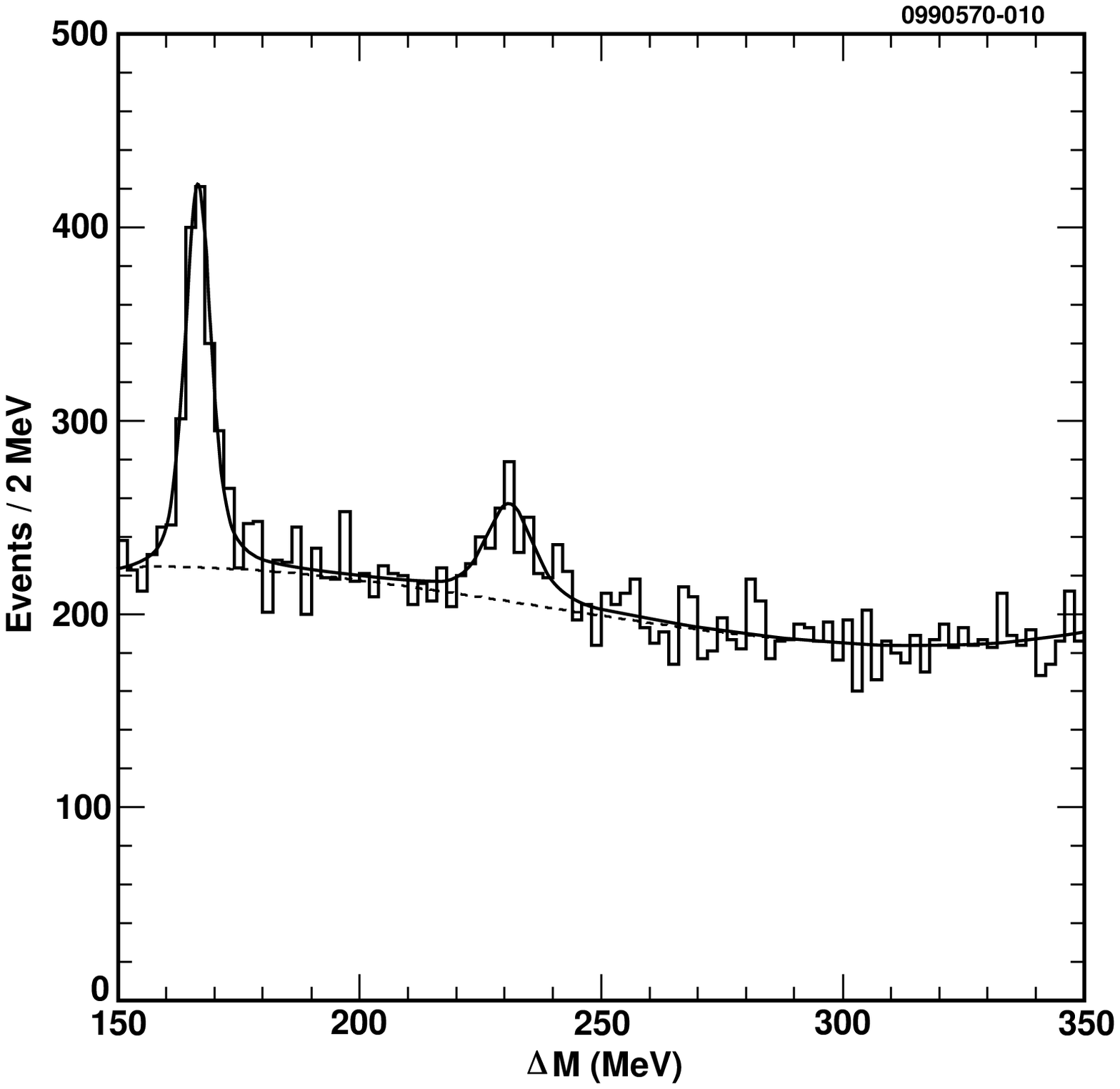}}
\caption{The distribution of $M(\Lambda_c \pi^0) - M(\Lambda_c)$
         showing the narrow $\Sigma_c^+$ and wide $\Sigma_c^{\ast +}$.}
\label{fig:sigma}
\end{figure}

	There are a great many results on charm baryons, and I can 
only cover the very highest points, and even these only briefly.
We have a first observation of the $\Sigma_c^{\ast +}$
and a new measure of $\Sigma_c^+$ mass both in
$\Lambda_c \pi^0$ decay.\cite{sigma}
The data is shown in Figure~\ref{fig:sigma} showing the clear
signals.

 	We make the first observation of
the $J^P = 1/2^-$ pair $\Xi^+_{c1}$ and
$\Xi^0_{c1}$.\cite{Xi}  Note that this result combined with the
observation of the $\Sigma_c^{\ast +}$ discussed above means
that all of the
$L=0$ single charm baryons have been observed.  The majority
of these have had their first observation by CLEO.

\begin{figure}
\resizebox{6in}{!}
{\includegraphics{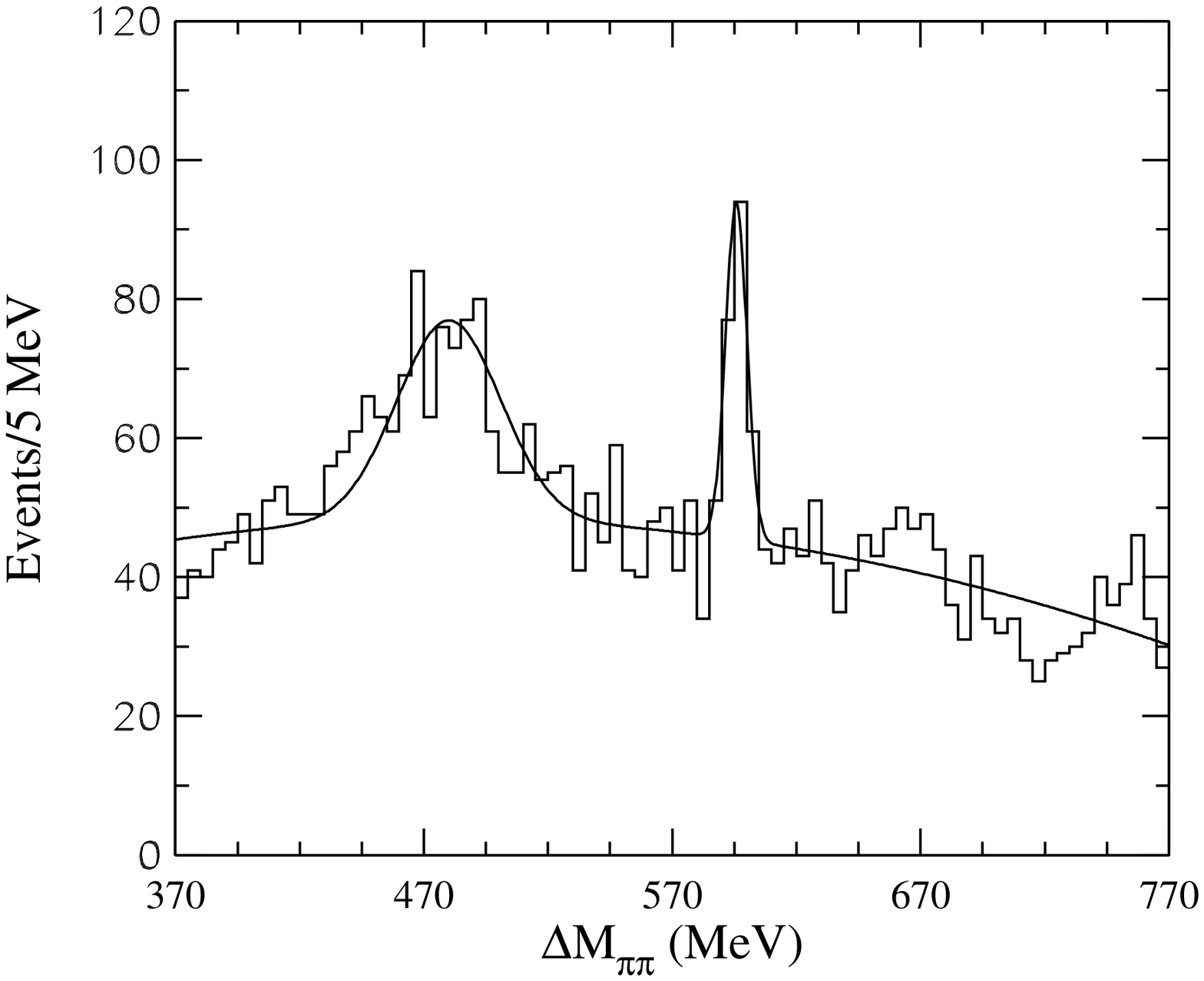}}
\caption{The distribution of $M(\Lambda_c \pi^+ \pi^-) - M(\Lambda_c)$
         showing the two new states.}
\label{fig:two}
\end{figure}

	In a preliminary result 
we see two states in decays to $\Lambda_c\pi^+\pi^-$;
a wide state decaying to $\Sigma_c$ and $\Sigma_c^\ast$, possibly
an orbitally excited $\Sigma_{c1}$, and a narrow state
decaying to $\Sigma_{c}\pi$ and non-resonant, possibly an
excited baryon with $L=1$ between the light quarks.\cite{two}  The data is
shown in Figure~\ref{fig:two}.  Our interpretation of these
two states is guided by the observed mass and width rather than
any observation of the angular distribution of their decays.

	CLEO has observed the $\Omega_c$ and made a precision measure
of its mass,\cite{omegac} and we have a new measurement of
the $\Lambda_c$ branching fraction into $p K \pi$.\cite{lambdac}

\section{Other Physics Results}
\subsection{$\Upsilon$(4S) Charged and Neutral Decay Fraction}

From the relative rates for $B^0 \to J/\psi K^{(\ast)0}$ and
$B^+ \to J/\psi K^{(\ast)+}$ we are able to measure the
relative branching fraction of the $\Upsilon(4S)$ to charged
and neutral $B$ mesons.\cite{f+-/f00} 
We find
\begin{eqnarray}
\frac{f_{+-}}{f_{00}} & \equiv \frac{\Gamma(\Upsilon(4S) \to B^+ B^-)}{\Gamma(\Upsilon(4S) \to B^0 \bar{B^0})} \\ \nonumber
  & = 1.04 \pm 0.07 \pm 0.04
\end{eqnarray}
and assuming $f_{+-} + f_{00} = 1$, that is the $\Upsilon(4S)$ only decays
to $B \bar B$, $f_{+-} = 0.49 \pm 0.02 \pm 0.01$
and $f_{00} = 0.51 \pm 0.02 \pm 0.014$.

\subsection{$\eta_c$ in $2\gamma$}

	We have observed 300 events of the two photon production
of the $\eta_c$.\cite{etac2g}  We measure 
$m(\eta_c) = (2980.4 \pm 2.3 \pm 0.6)$ MeV, 
$\Gamma(\eta_c) = (27.0 \pm 5.8 \pm 1.4)$ MeV, and the two photon
partial width $\Gamma_{2\gamma}(\eta_c) = (7.6 \pm 0.8 \pm 0.4 \pm 2.3)$ keV,
where the last error is from the branching fraction
of the $\eta_c$ into $K_s^0 K^\mp \pi^\pm$.  This last observation
agrees much better with the prediction of Peturbative QCD based
on the $e^+e^-$ partial width of the $J/\psi$.

\section{CLEO~III Status}

	The CLEO~II.V detector ceased data taking in February of 1999.
It has been upgraded to the CLEO~III detector.  The upgrade consists
of a new four layer double sided silicon drift detector, a new
47 layer drift chamber, much stronger particle ID with the addition
of a new barrel RICH, refurbished CsI
calorimeter and muon system,
and a new data acquisition and trigger system that are designed
to handle delivered luminosity of up to
$5 \times 10^{33}/{\rm cm}^2\!-\!{\rm sec}$.  The upgraded detector was completed
in April of 2000 and started taking physics data in July of 2000.

	The new detector is performing well with a tracking system already
working as well as the CLEO~II configuration, and a much improved
resolution on photons in the calorimeter endcaps due to the reduction
of support material in the upgrade from the old to the new drift
chamber.  The RICH
is also performing very well with a preliminary efficiency of around
90\%, and an intrinsic resolution on the Cherenkov angle of 2-4 mrad. 

\section{Conclusion}

	There is a vast array of physics results to be had from the
$\Upsilon(4S)$.  Highlights from CLEO in the last six months are
an unambiguous observation of the gluonic penguin, the best single
measure of $|V_{cb}|$, and the first observation of a high multiplicity
hadronic $B$ decay mode.  There are many other results and with the
beginning of CLEO~III data taking and first results from the asymmetric
$B$ factories we can all look forward to much more exciting physics
from the $\Upsilon(4S)$ in the future.

\end{document}